\documentclass[10pt,aps,prb,twocolumn,superscriptaddress]{revtex4}
\usepackage{amsmath}
\usepackage{amsfonts}
\usepackage{amssymb}
\usepackage{graphicx}
\usepackage{bm}
\usepackage{float}
\usepackage{appendix}

\def\bsub{\begin{subequations}}
\def\esub{\end{subequations}}

\begin{document}
\title{Witnessing a Poincar\'e recurrence with Mathematica}
\author{J.~M.~Zhang}
\email{wdlang06@163.com}
\affiliation{Fujian Provincial Key Laboratory of Quantum Manipulation and New Energy Materials,
College of Physics and Energy, Fujian Normal University, Fuzhou 350007, China}
\affiliation{Fujian Provincial Collaborative Innovation Center for Optoelectronic Semiconductors and Efficient Devices, Xiamen, 361005, China}


\author{Y. Liu}
\email{liu_yu@iapcm.ac.cn}
\affiliation{Institute of Applied Physics and Computational Mathematics, Beijing 100088, China}
\affiliation{Software Center for High Performance Numerical Simulation, China Academy of Engineering Physics, Beijing 100088, China}

\begin{abstract}
The often elusive Poincar\'e recurrence can be witnessed in a completely integrable system. For such systems, the problem of recurrence reduces to the classic mathematical problem of simultaneous Diophantine approximation of multiple numbers. The latter problem then can be somewhat satisfactorily solved by using the famous Lenstra-Lenstra-Lov\'{a}sz (LLL) algorithm, which is implemented in the Mathematica built-in function \verb"LatticeReduce". The procedure is illustrated with a harmonic chain. The incredibly large recurrence times are obtained exactly. They follow the expected scaling law very well.
\end{abstract}

\maketitle

\section{Introduction}

The theorem established by Poincar\'e in 1890 is both stunning and intriguing.\cite{poincare} It states that for any Hamiltonian system with a finite accessible phase space, the system will return to the proximity of a generic initial state infinitely many times. For instance, it would assert that if  a cloud of gas initially confined in the left compartment of a vessel is released into the right empty compartment, then after a sufficiently long time, the gas molecules will reassemble in the left compartment.
Moreover, they will do so again and again indefinitely.

The power of this paradoxical theorem lies in its generality. It results from the volume-preserving property (Liouville's theorem\cite{landau}) of the Hamiltonian flow and the finiteness\cite{non} of the phase space. The former is a common to any  Hamiltonian system and the latter can be easily satisfied in many cases, say, the cloud of molecules in the scenario above. A key insight of Poincar\'{e} is that, because of the volume-preserving property, for any initial region $V_0 $, its time-evolved copies at times $0$, $\tau$, $2 \tau$, $\ldots$, denoted respectively as $V_0$, $V_1$, $V_2$, $\ldots$, are all of the same volume as $V_0$, and since the total accessible phase space has a finite volume, two of these regions must overlap. Let $V_m \bigcap V_n \neq 0 $ ($m<n $), then we have $V_0 \bigcap V_{n-m } \neq 0 $. That is, some points in $V_0 $ come back into $V_0 $ at $(n-m )\tau $. Complete proofs and the rigorous statements can be found in the original paper by Poincar\'e and many books on dynamical systems and/or ergodic theory.\cite{walters}

While the proof is not complicated and the conclusion inescapable, the predicted recurrence still seems elusive.  The problem is that the proof is not constructive---It merely guarantees the existence of the recurrence, but does not tell us when it will occur. In a reply to Zermelo, who used the Poincar\'{e} recurrence to question the validity of the $H$-theorem,\cite{zermelo,steckline} Boltzmann argued that,\cite{boltzmann} as is widely accepted today, the recurrence time is extremely large. But how long exactly is it? Can we predict a Poincar\'e recurrence as we can predict the recurrence of Halley's comet? Although this is a question out of curiosity, it is not without pedagogical value. Usually, the recurrence time is estimated rudely, with the calculation based more on probability than on mechanical considerations. It is definitely more persuasive to do a rigorous calculation respecting all the mechanical laws and get the number exactly.

The task is apparently a difficult one for a generic system. However, as shown in previous works, for a special class of systems, completely integrable systems to be concrete, the problem is tractable. For such systems, the formidable problem of integrating the equation of motion can be bypassed, and the recurrence problem can be reduced to the classic problem of simultaneous Diophantine approximation of real numbers.\cite{frisch,hemmer,kim}  For this classic problem in number theory, there are tons of deep and beautiful theorems,\cite{cassels} and on the computational side, there is the celebrated Lenstra-Lenstra-Lov\'{a}sz (LLL) algorithm.\cite{LLL} Fortunately for our purpose, this algorithm is implemented in Mathematica in the function \verb"LatticeReduce". By invoking this function, we can obtain effortlessly the astronomically large recurrence times exactly.

\begin{figure}[tb]
\includegraphics[width= 0.4\textwidth ]{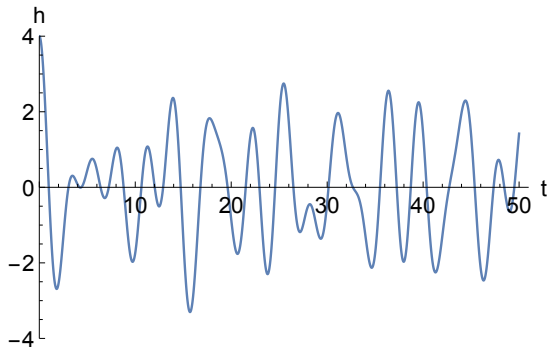}
\caption{The $h$ function defined in (\ref{hfunction}). Its initial value at $t= 0 $ is 4, which is also the maximal value it can achieve. Because of dephasing of the different cosine terms, $h $ can hardly return close to its initial value for a long time.   }
\label{hplot}
\end{figure}

Before proceeding, let us recall how the paper was motivated. In 2014, when the first author was a postdoc in the Max Planck Institute for the Physics of Complex Systems, out of curiosity he offered 100 Euro for a number $t$, such that $t>10$  and the function (see Fig.~\ref{hplot})
\begin{eqnarray}\label{hfunction}
  h(t) = \cos(t) + \cos(\sqrt{2}t) + \cos(\sqrt{3}t) + \cos(\sqrt{5} t)
\end{eqnarray}
is close to $4$ within $10^{-6}$, i.e., $ |h(t) -4|  \leq 10^{-6}$. The question was posed because of its apparent relevance to the Poincar\'{e} recurrence. For $h$ to be close to $4$, each cosine term should be close to unity, or, the four numbers $\{t,\sqrt{2}t,\sqrt{3}t,\sqrt{5}t \}$ should be close to integer multiples of $2\pi $ simultaneously, a condition essentially the same as that of Poincar\'{e} recurrence of a completely integrable system.  By brute-force, a colleague found the huge number
\begin{eqnarray}\label{num1}
  t &=& 2\pi\times  10\,458\,943\,416 ,
\end{eqnarray}
and won the money. But the brute-force method apparently is out of the question if there are more cosine terms or if a much higher precision is required, and a more general and sophisticated approach was really what was wanted. Later, we learnt of the LLL algorithm.

\section{Poincar\'{e} recurrence of completely integrable systems}

For a generic multi-degree-of-freedom system, its dynamics is chaotic, hence it is hard to follow its motion for a long time, let alone to predict the Poincar\'e recurrences in the extremely distant future.

More trackable is the motion of completely integrable systems.\cite{landau} For such a system executing a motion finite in all coordinates,  the picture is very simple in terms of the action variables $J_i $ and angle variables $\theta_i $ ($1\leq i \leq N  $). The Hamiltonian
\begin{eqnarray}
  H = H(J_1, \ldots, J_{N})
\end{eqnarray}
depends only on the action variables, hence all the action variables are conserved, while all the angle variables increase linearly with time,
\begin{eqnarray}\label{eoqangle}
  \dot{\theta}_i  = \frac{\partial H }{\partial J_i } \equiv \omega_i .
\end{eqnarray}
As the angle variables are defined modulo $2 \pi$, the system evolves on a \emph{torus}, the radiuses of which are determined by the actions and the surface of which is parametrized by the angle variables. Explicitly,  the torus is
\begin{eqnarray}
  \mathbb{T}^{N} = \{ (\theta_1, \ldots, \theta_{N} )| \theta_i\in[0 , 2 \pi), 1\leq i \leq N  \}  .
\end{eqnarray}
The flow defined in (\ref{eoqangle}) is volume-preserving and the total volume of the torus $\mathbb{T}^{N }$ is finite, hence the two prerequisites for Poincar\'{e} recurrence are satisfied. For a recurrence, all the angle variables should return close to their initial values. In particular, the variable $\theta_{N} $ should return close to its initial value. Since it has a period of $T = 2\pi /\omega_{N} $, we take the snapshots of the continuous evolution defined by (\ref{eoqangle}) with a period of $T$, and consider the resulting discrete dynamical system defined on the torus
\begin{eqnarray}
  \mathbb{T}^{N-1}  =\{ (\theta_1, \ldots, \theta_{N-1} )| \theta_i \in [0,  2 \pi ), 1\leq i \leq N-1 \} ,
\end{eqnarray}
namely,
\begin{eqnarray}\label{f}
  f(\theta_1, \ldots, \theta_{N-1}) = (\theta_1 + 2\pi \alpha_1, \ldots ,\theta_{N-1} + 2\pi \alpha_{N-1}) ,
\end{eqnarray}
where the variables $\alpha_i $ are defined as ($1\leq i \leq N-1 $)
\begin{eqnarray}
 \alpha_i \equiv  \frac{\omega_i }{\omega_{N}}  .
\end{eqnarray}
Apparently, $f$ is also volume-preserving.

Now take a small neighborhood of the origin
\begin{eqnarray}
 \Omega = \{ (\theta_1, \ldots, \theta_{N-1})| |\theta_i | \leq  \pi \epsilon  \}
\end{eqnarray}
with $\epsilon \ll 1 $. Consider the images of $\Omega $ under the iterated action of $f$, i.e., $\Omega_i = f^i (\Omega)$, $0\leq i \leq \infty $. They are all of the same volume. Hence, for $K  = \lfloor 1/\epsilon^{N-1} \rfloor $, where $\lfloor x \rfloor $ denotes the largest integer no more than $x$, the $K +1 $ regions $\{\Omega_i | 0\leq i \leq K \}$ cannot be mutually disjoint. Otherwise,
\begin{eqnarray}
  \text{Vol}(  \bigcup_{i=0}^K \Omega_i ) &=& (K+1) (2\pi \epsilon )^{N-1} \nonumber \\
  & > & (2\pi)^{N-1} = \text{Vol}(\mathbb{T}^{N-1}).
\end{eqnarray}
Hence there must be $0\leq m < n \leq K $ such that
\begin{eqnarray}\label{overlap1}
\Omega_m \bigcap \Omega_n \neq 0 .
\end{eqnarray}
Now by applying $f^{-m }$ to both sides, we get ($q = n - m \leq K $)
\begin{eqnarray}\label{overlap2}
\Omega_0 \bigcap \Omega_{q } \neq 0.
\end{eqnarray}
This means recurrence. Specifically, let $ (\tilde{\theta}_1, \ldots,  \tilde{\theta}_{N-1}) \in  \Omega_0 \bigcap \Omega_{q } $. Since this point belongs to $\Omega_q $, there exists a point  $(\theta_1, \ldots, \theta_{N-1})\in \Omega_0 $ which,  when translated $q$ times by $f$, arrives at $(\tilde{\theta}_1, \ldots,  \tilde{\theta}_{N-1} ) \in \Omega_0 $, i.e., it comes back into $\Omega_0$. The relative error of recurrence is bounded by
\begin{eqnarray}
   \max_{1\leq i \leq N - 1 }   |\theta_i -\tilde{\theta}_i|   \leq  2 \pi \epsilon ,
\end{eqnarray}
 as $\{|\theta_i | , |\tilde{\theta_i }|\}\leq \pi \epsilon \ll \pi $.
 As $f$ translates the torus as a whole rigidly, when one point returns to the proximity of its initial position, so do all other points. That is, the system not only recurs, but  even recurs uniformly, or in other words, independent of the initial condition. We have thus not only proved the existence of recurrence, but also obtained an upper bound of the recurrence time in terms of the precision $\epsilon$ and the number of degrees of freedom. The scaling law anticipated is that the recurrence time is on the order of $1/\epsilon^{N-1}$ for a completely integrable system with $N$ degrees of freedom.

It is easily seen that the condition (\ref{overlap2}) is equivalent to the condition that
\begin{eqnarray}\label{diophantine1}
  \langle q \alpha_i \rangle \equiv  |q \alpha_i - \lfloor q \alpha_i \rceil |  \leq  \epsilon
\end{eqnarray}
for each $1\leq i \leq N - 1 $. Here, $\lfloor x \rceil $ denotes the integer nearest to the real number $x$ (for instance, $\lfloor 2.4 \rceil = 2$ and $\lfloor 2.7\rceil = 3$), and $\langle x \rangle  $ denotes the distance of $x$  to $\lfloor x \rceil $. By definition, $\langle x \rangle \leq 1/2$ for an arbitrary $x$.

Equation (\ref{diophantine1}) is the sufficient condition for a recurrence with a prescribed precision of $\epsilon $. The problem of looking for a Poincar\'{e} recurrence is then reduced to the problem of looking for an integer $q$ such that (\ref{diophantine1}) is satisfied simultaneously for all the numbers $\{\alpha_1, \ldots, \alpha_{N-1} \}$. It turns out that this is a classic problem termed simultaneous Diophantine approximation in number theory.\cite{cassels} For this problem, there is the celebrated Dirichlet theorem, which states that for $n$ arbitrary real numbers $\{\alpha_1, \ldots, \alpha_n \}$, and any $\epsilon > 0$, there exists a positive integer $q \leq 1/\epsilon^n$, such that
\begin{eqnarray}\label{diophantine2}
  \max \{\langle q \alpha_1 \rangle, \ldots, \langle q \alpha_n \rangle   \} \leq  \epsilon.
\end{eqnarray}
This is essentially the same as what is proven above.

\section{LLL algorithm}
\begin{figure}[tb]
\includegraphics[width= 0.4\textwidth ]{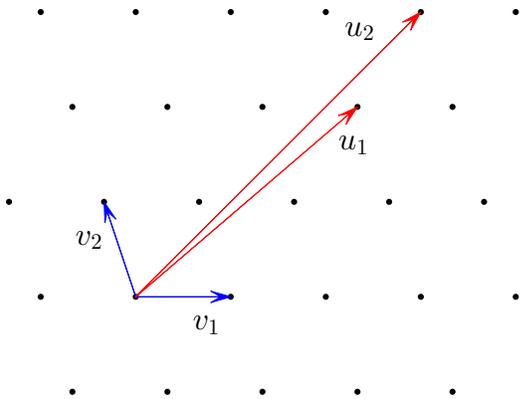}
\caption{A two-dimensional lattice and two of its  bases. Although both the $u$-basis $\{u_1, u_2\}$ and the $v$-basis $\{v_1, v_2 \}$ span the lattice, the latter is reduced comparatively, in the sense that the basis vectors are shorter and are more close to being orthogonal. }
\label{lattice2d}
\end{figure}

The problem is now to find a $q$ such that (\ref{diophantine1}) is satisfied for given $\epsilon $, which will be arbitrarily small. For this purpose, a useful tool is the famous LLL algorithm, which is implemented in Mathematica by the function \verb"LatticeReduce". As the name suggests, the algorithm is about the lattice, a most basic concept in solid state physics. Suppose $\{ \vec{u}_1, \ldots, \vec{u}_n \}$ are $n $ linearly independent vectors in $\mathbb{R}^n$. The lattice determined by them is the subset of $\mathbb{R}^n $ defined as
\begin{eqnarray}
  L &=& \{\sum_{i=1}^n r_i \vec{u}_i | r_i \in \mathbb{Z} \} .
\end{eqnarray}
The vectors $\{ \vec{u}_1, \ldots, \vec{u}_n \} $ are called a basis of the lattice. We say they span the lattice.
The point is that a basis determines a lattice, but not vice versa. In other words, for a given lattice, the basis is not unique.
Actually, for any $n\times n $ matrix $M $, whose entries are all integral and whose determinant is $\pm 1 $, the vectors
\begin{eqnarray}\label{transform}
  \vec{v}_i  &=& \sum_{j=1}^n M_{ij } \vec{u}_j
\end{eqnarray}
constitute another basis of $L $. The proof is simple. By (\ref{transform}), the lattice spanned by the $\vec{v}$'s, denoted as $L'$, is contained in $L$. However, by construction, the inverse of $M $ has the same property as $M$, and hence the $\vec{u}$'s are also integral linear combinations of the $\vec{v}$'s, which means that $L$ is contained in $L'$. Therefore, they must be equal.

For a lattice, a characteristic quantity is its determinant, which is defined as
\begin{eqnarray}
  d(L) &=& |\det  (\mathcal{U })| .
\end{eqnarray}
Here $\mathcal{U} \equiv (\vec{u}_1 ; \ldots ; \vec{u}_n )$ is the $n \times n $ matrix with $\vec{u}_i $ being the $i$th row. Despite of the explicit reference to the basis in its definition, the quantity $d(L )$ does not depend on the choice of the basis actually. The reason is simple---For another basis $\{ \vec{v}_i \}$ of $L $, the matrix $\mathcal{V} \equiv (\vec{v}_1 ; \ldots ; \vec{v}_n ) $ relates to the matrix $\mathcal{U}$ as $\mathcal{V} = M \mathcal{U}$ by (\ref{transform}), and therefore,
\begin{eqnarray}
  |\det(\mathcal{V})| = |\det(M ) \det(\mathcal{U })| = |\det(\mathcal{U })| .
\end{eqnarray}
The physical or geometric meaning of $d(L)$ is the volume of the parallelepiped subtended by the $n $ basis vectors. Hence, we have the Hadamard's inequality\cite{matrix}
\begin{eqnarray}\label{hadamard}
  d(L) \leq  \prod_{i=1}^n  |\vec{u}_i |    .
\end{eqnarray}
The equality is achieved if and only if $\vec{u}_i $ are orthogonal to each other.

Although all basis are equivalent in spanning the lattice, some basis is more preferable than others. For example, in Fig.~\ref{lattice2d}, the lattice is spanned either by the basis
\begin{eqnarray}
  B_1  &=& \{ \vec{u}_1 = (\frac{7}{3}, 2), \vec{u}_2 = (3,3) \},
\end{eqnarray}
or by the basis
\begin{eqnarray}
  B_2  &=& \{ \vec{v}_1 = (1, 0), \vec{v}_2 = (-\frac{1}{3} ,1) \}.
\end{eqnarray}
Apparently, the latter is more appealing as the basis vectors involved are shorter and more close to being orthogonal. We thus see that the defining basis of a lattice might be far from being optimal and generally there is room to reduce or simplify it, in the sense that $B_2 $ is reduced compared to $B_1 $. This is exactly what the LLL algorithm is about.

There are numerous literature about the LLL algorithm. In particular, the original paper is very readable.\cite{LLL} However, here for our purpose, we just need to know that the LLL algorithm, when fed with an original basis $\{ \vec{a}_1, \ldots, \vec{a}_n   \}$, will return us in polynomial time a new basis $\{\vec{b}_1, \ldots, \vec{b}_n \}$ such that
\begin{eqnarray}\label{b1}
  |\vec{b}_1 | \leq 2^{(n-1)/4} [d(L )]^{1/n }  .
\end{eqnarray}
That the new basis is somehow reduced can be glimpsed by comparing (\ref{b1}) with (\ref{hadamard}). The determinant $d (L)$ and the basis vector lengths swapped position.

The way the LLL algorithm is used for simultaneous Diophantine approximation is as follows. Let $\{ \alpha_1, \ldots, \alpha_n \}$ be the $ n $ real numbers. Construct the $(n+1)\times (n+1 )$ matrix
\begin{eqnarray}\label{amatrix}
  A &=& \left(
          \begin{array}{cccc}
            1 & Q \alpha_1   & \cdots  & Q \alpha_n   \\
             & Q &  &  \\
             &  & \ddots &  \\
             &  &  & Q \\
          \end{array}
        \right),
\end{eqnarray}
where $Q $ is some large integer to be chosen according to the precision wanted. The $(n+1)$ rows are the basis of a lattice. The determinant of the lattice is apparently $d(L) = Q^n $.  The first vector $\vec{b}_1 $ in the reduced basis is of the form
\begin{eqnarray}
  \vec{b}_1 &=& (q, Q (q \alpha_1 - p_1 ), \ldots, Q (q \alpha_n - p_n )),
\end{eqnarray}
where $q$ and $p_i $ ($1\leq i \leq n $) are integers. By (\ref{b1}), we have
\bsub
\begin{eqnarray}
  q \leq |\vec{b}_1| \leq 2^{n/4} Q ^{n/(n+1)} ,  \\
  Q |q \alpha_i - p_i |  \leq   |\vec{b}_1| \leq 2^{n/4} Q ^{n/(n+1)} .
\end{eqnarray}
\esub
Let $Q =  2^{n( n+1)/4} C^{n+1}$, we have
\bsub\label{limit1}
\begin{eqnarray}
  q &\leq & 2^{n (n+1 ) /4} C ^{n} ,  \\
 |q \alpha_i - p_i | & \leq &  C^{-1} .
\end{eqnarray}
\esub
Hence, we have found a $q$ such that (\ref{diophantine2}) is satisfied with $\epsilon = 1/ C $. This $q$ is not bounded by $1/\epsilon^n $ but is on the same order.

In practice, the LLL algorithm is generally implemented with the components of the basis vectors being rational numbers. Hence, a small modification is necessary.\cite{hanrot} Instead of (\ref{amatrix}), we construct the matrix
\begin{eqnarray}\label{bmatrix}
  B &=& \left(
          \begin{array}{cccc}
            1 & \lfloor Q  \alpha_1 \rceil   & \cdots  &  \lfloor  Q \alpha_n \rceil   \\
             & Q &  &  \\
             &   & \ddots &  \\
              &   &  & Q \\
          \end{array}
        \right),
\end{eqnarray}
where the rounding function $\lfloor\cdot \rceil $ is implemented in Mathematica with \verb"Round".
In this case,
\begin{eqnarray}
  \vec{b}_1 &=& (q, q \lfloor Q \alpha_1 \rceil - p_1 Q  , \ldots, q \lfloor Q \alpha_n \rceil - p_n Q).
\end{eqnarray}
By (\ref{b1}), we have
\begin{eqnarray}\label{q1}
  q \leq |\vec{b}_1 | \leq 2^{n/4} Q^{n/(n+1)}.
\end{eqnarray}
As for $|q \alpha_i - p_i |$, we have
\begin{eqnarray}\label{error}
  |q \alpha_i - p_i | &= & \frac{1}{Q } |q Q \alpha_i - p_i Q  | \nonumber \\
   &\leq & \frac{1}{Q} (| q  \lfloor Q \alpha_i \rceil - p_i Q | + \frac{q}{2} )  \nonumber  \\
   &\leq &  \frac{1}{Q} \sqrt{1 + \frac{1}{4}} \sqrt{( q  \lfloor Q \alpha_i \rceil - p_i Q )^2 + q^2 } \nonumber \\
   & \leq & \frac{1}{Q} \frac{\sqrt{5}}{2} |\vec{b}_1 | \leq  \frac{\sqrt{5}}{2}  2^{n/4} Q^{-1/(n+1)}.
\end{eqnarray}
Here in the second line, we used the inequality $ |\alpha_i- \lfloor \alpha_i \rceil | \leq 1/2  $, and in the third line, we used the Cauchy-Schwarz inequality. Taking $Q = (\frac{\sqrt{5}}{2})^{n+1} 2^{n(n+1)/4} C^{n+1}$, we have similar to (\ref{limit1}),
\bsub\label{limit2}
\begin{eqnarray}
  q  &\leq & 5^{n/2} 2^{n(n-3)/4} C^{n} , \\
  |q \alpha_i - p_i | &\leq &    C^{-1}.
\end{eqnarray}
\esub
The scaling law is unchanged. This is the very algorithm we shall use to search the exact recurrence times.

\section{A case study}
\begin{figure}[tb]
\includegraphics[width= 0.4\textwidth ]{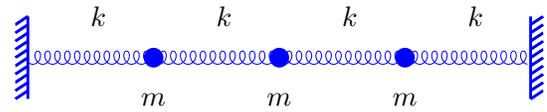}
\caption{The one-dimensional harmonic chain. Each of the $N $ particles is of mass $m$. Two adjacent particles are linked by a spring of stiffness $k$. Here for illustration we have $N = 3$.   }
\label{figchain}
\end{figure}

\begin{figure*}[tb]
\includegraphics[width= 0.8\textwidth ]{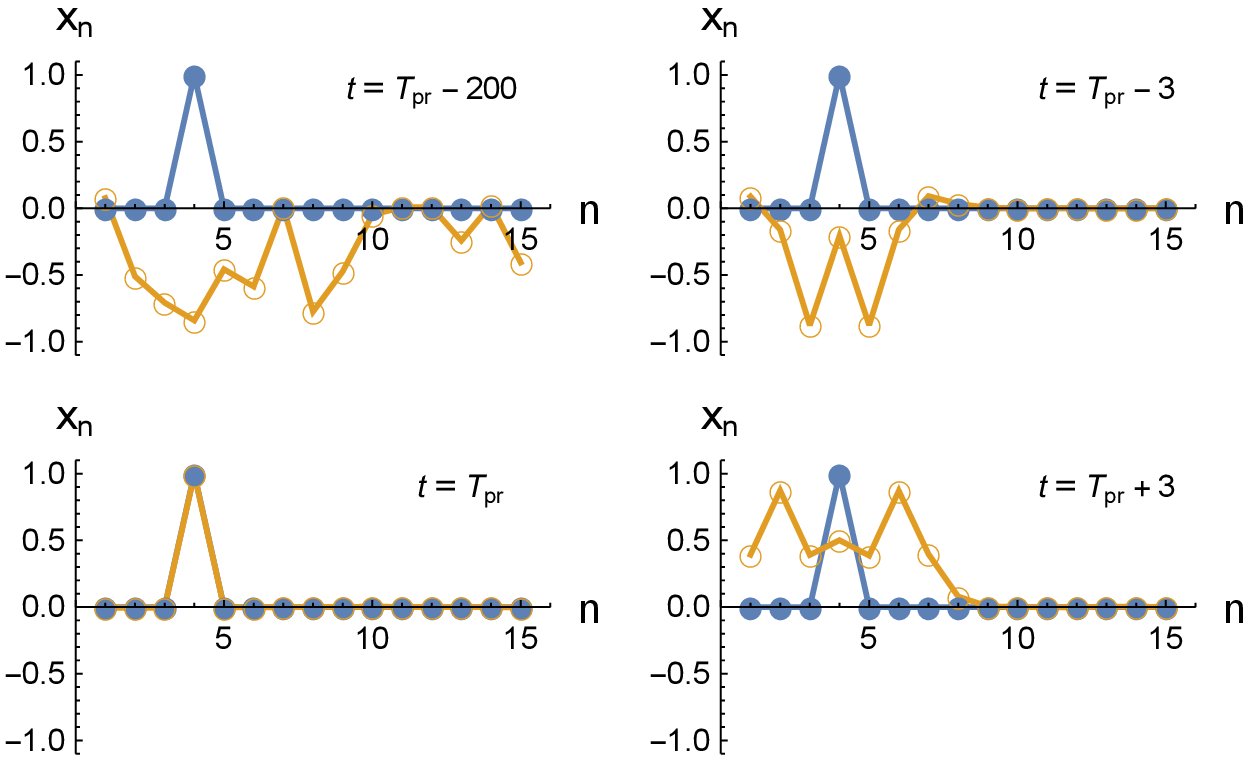}\\
\vspace{0.5cm}
\includegraphics[width= 0.8\textwidth ]{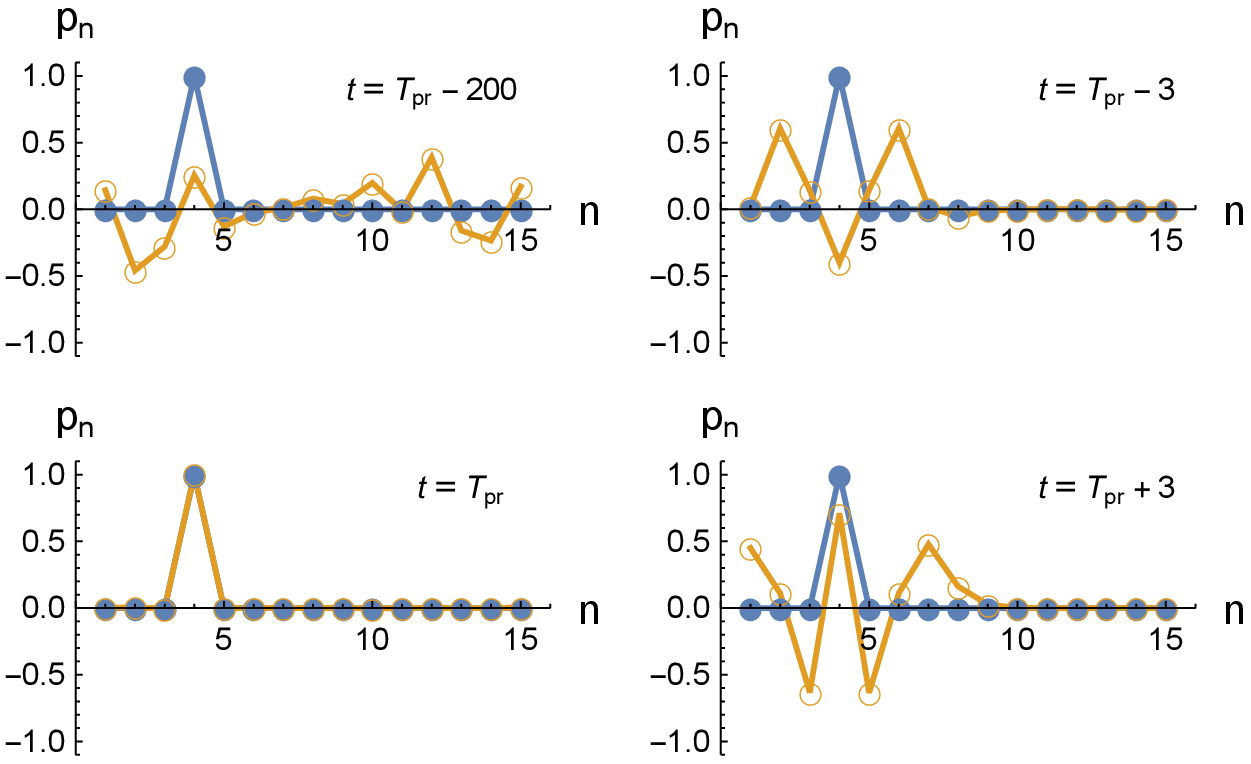}
\caption{Coordinates $x_n$ and momenta $p_n $ of the point-masses at $t = T_{pr}-200$, $t = T_{pr} $, and $t = T_{pr} \pm 3 $. Here $T_{pr} $ is the Poincar\'{e} recurrence time defined in (\ref{defprt}) and (\ref{particular}). In each panel, the $\bullet$ markers denote the initial state defined in (\ref{ini}) with $k=4$, while the $\circ$ markers denote the current state.  The recurrence of the initial state at $t = T_{pr}$ is to a very good extent as the $\circ$ markers coincide with the $\bullet $ markers almost completely. }
\label{snapshots}
\end{figure*}

We now demonstrate the theory described above with a concrete model, which is illustrated in Fig.~\ref{figchain}. The $N $ point-masses are each of mass $m $ and the springs are all of stiffness $k = m \omega^2   $ (we shall set $m=k=\omega = 1$ when it comes to numerics). The Hamiltonian is
\begin{eqnarray}\label{h}
  H   =  \frac{1}{2m }\mathbf{p}^T  \mathbf{p} + \frac{1}{2}m \omega^2 \mathbf{x}^T  \mathbf{M}  \mathbf{x} .
\end{eqnarray}
Here $\mathbf{x} \equiv (x_1, x_2, \ldots, x_N)^T$ denotes the displacements of the point-masses with respect to their equilibrium positions and $\mathbf{p} \equiv (p_1, p_2, \ldots, p_N )^T$ denotes their momenta. The $N\times N $ matrix $\mathbf{M }$ is tridiagonal as
\begin{eqnarray}
  \mathbf{M} &=&  \left(
               \begin{array}{ccccc}
                 2 & -1 &  &  & \\
                 -1 & 2 & \ddots  &  & \\
                 & \ddots & \ddots & \ddots & \\
                  &  & \ddots & 2 & -1  \\
                  &  &   & -1 & 2  \\
               \end{array}
             \right).
\end{eqnarray}
It is symmetric and thus can be diagonalized. That is, we have
\begin{eqnarray}\label{mdiag}
  \mathbf{M }= \mathbf{U} \mathbf{D } \mathbf{U}^T  ,
\end{eqnarray}
with $\mathbf{D } $ being a diagonal matrix and $\mathbf{U} $ an orthogonal matrix. The explicit forms of $\mathbf{D}$ and $\mathbf{U} $ are
\bsub\label{dandu}
\begin{eqnarray}
  D_{ij } &=&  4 \sin^2 \frac{j \pi }{2( N + 1 ) }  \delta_{ij } ,  \\
  U_{ij } &=&  \sqrt{\frac{2}{N+ 1}} \sin \frac{ij \pi }{N + 1} ,
\end{eqnarray}
\esub
with $1\leq i, j \leq N $. Substituting (\ref{mdiag}) and (\ref{dandu}) into (\ref{h}), we get
\begin{eqnarray}
  H  &=&  \frac{1}{2m }\mathbf{p}^T \mathbf{U}   \mathbf{U}^T  \mathbf{p} + \frac{1}{2}m \omega^2 \mathbf{x}^T \mathbf{U} \mathbf{D}  \mathbf{U}^T \mathbf{x} \nonumber \\
  &=& \frac{1}{2m }\mathbf{P}^T  \mathbf{P} + \frac{1}{2}m \omega^2 \mathbf{X}^T  \mathbf{D}  \mathbf{X } \nonumber\\
  &=& \sum_{i=1}^N \left( \frac{P_i^2}{2m} + \frac{1}{2}m \omega_i^2 X_i^2   \right).
\end{eqnarray}
Here we have introduced the new coordinates $\mathbf{X} \equiv (X_1, X_2,
\ldots, X_N )^T$ and momenta $  \mathbf{P} \equiv (P_1, P_2, \ldots, P_N )^T  $,
\begin{eqnarray}
  (\mathbf{X}, \mathbf{P}) \equiv  \mathbf{U}^T (\mathbf{x}, \mathbf{p}).
\end{eqnarray}
The Hamiltonian is thus diagonalized as the sum of $N $ independent harmonic oscillators, with the frequency of the $i$th oscillator being
\begin{eqnarray}\label{frequency}
  \omega_i  = 2 \omega \sin \frac{i \pi }{2(N + 1) } .
\end{eqnarray}
The evolution of the system is then  simply that
\begin{eqnarray}
  (P_i(t) ,X_i (t ) ) = A_i (\cos (\omega_i t + \phi_i ), \sin (\omega_i t + \phi_i ) ),\;
\end{eqnarray}
where the amplitude $A_i $ and the phase $\phi_i $ are determined by the initial condition. We thus have the $N$-torus formalism and the algorithm can be applied directly.

The procedure to get a recurrence time is then as follows. Let $\alpha_i = \omega_i /\omega_N $, $1\leq i \leq N -1 $. Choose a large integer $Q $, and construct the $B$ matrix as in (\ref{bmatrix}). Feed it into the function \verb"LatticeReduce", and get a reduced basis. The corresponding code is
\begin{verbatim}
   RB = LatticeReduce(B);
\end{verbatim}
The new basis vectors are the rows of \verb"RB", and the integer $q$ we want is the first component of the first basis vector, i.e.,
\begin{verbatim}
   q = RB[[1,1]];
\end{verbatim}
Once the $q$ is obtained, the recurrence time is
\begin{eqnarray}\label{defprt}
  T_\text{pr}  = \frac{2\pi q}{\omega_N }  .
\end{eqnarray}
Here the subscript means Poincar\'{e} recurrence.

How good the algorithm works can be checked with some concrete numbers. Let $ N = 15$, and $Q = 10^{35}$, we get
\begin{eqnarray}\label{particular}
  q = 84\,350\,294\,911\,456\,044\,599\,486\,768\,675\,168,
\end{eqnarray}
and the recurrence error
\begin{eqnarray}\label{recerror}
 error =  \max_{1\leq i \leq N-1 }  \langle q \alpha_i \rangle   = 0.002722.
\end{eqnarray}
The error is sufficiently small, hence an arbitrary initial state will recur after time $T_\text{pr} = 2\pi q /\omega_N $ to a good precision. Consider the initial state defined as
\begin{eqnarray}\label{ini}
x_i=p_i  = \begin{cases} 0, & 1\leq i \leq k - 1, \\ 1, & i = k , \\ 0, &  k+1 \leq i \leq N .  \end{cases}
\end{eqnarray}
Note that we have deliberately chosen an artificial state, in which only the $k$th point-mass is displaced and has a nonzero velocity. In Fig.~\ref{snapshots}, four snapshots of the coordinates and momenta of the point-masses are shown, with the corresponding times being $t = T_\text{pr} -200 $, $t = T_\text{pr}$, and $t = T_\text{pr} \pm 3 $. We see that at $t = T_\text{pr}-200 $, almost all the point-masses are significantly displaced and have gained significant momenta. This is typical of the configuration of the system at an arbitrary time. The initially localized motion has spread over the whole system and the system remains so for a long time. However, around $t = T_\text{pr}$, the system exhibits a recurrence. At $t = T_\text{pr} - 3$, the tendency is clear, the motion regathers around the $k$th point-mass, and the point-masses far away have already returned to their initial positions and come to rest. Finally, at $t =T_\text{pr}$, the system restores itself to its initial configuration to such good extent that the difference between the current values  and the initial values of $(x_n,  p_n )$ is hardly visible. Afterwards, as shown by the snapshots at $t = T_\text{pr} +3 $, the motion disperses again and we have to wait yet another long time to witness a second recurrence.

\begin{figure*}[tb]
\includegraphics[width= 0.45\textwidth ]{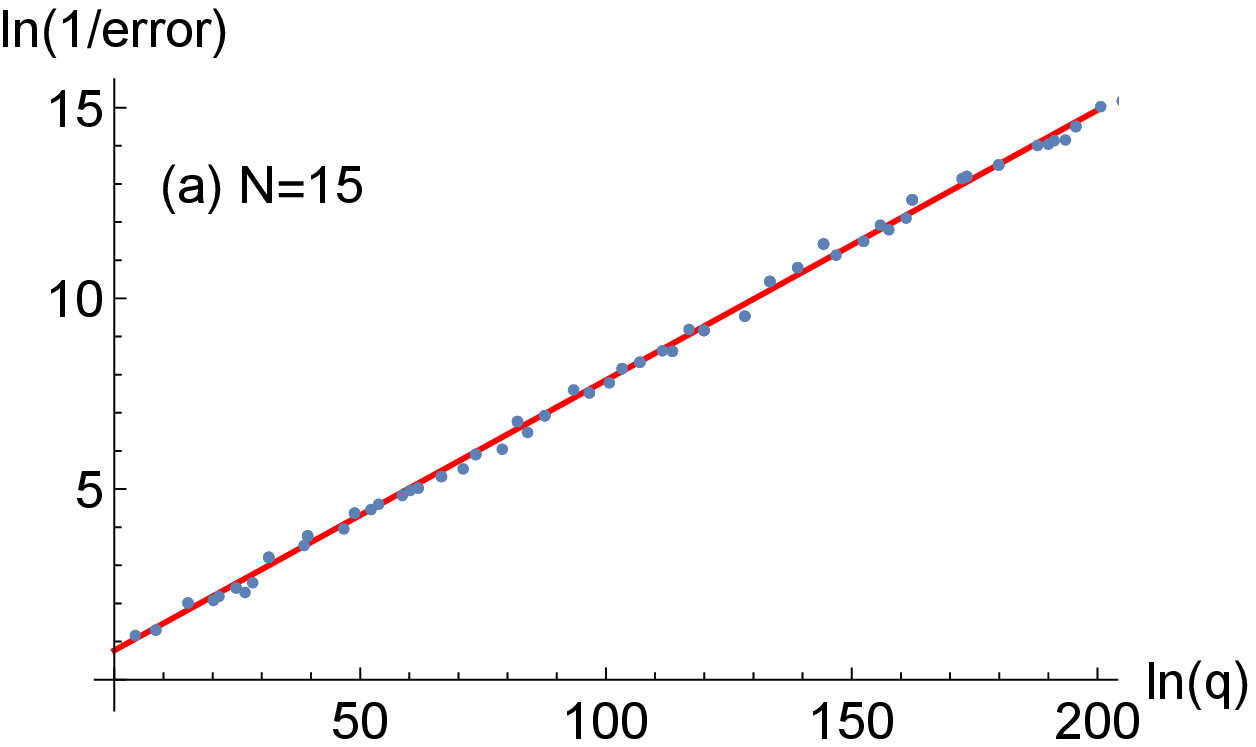}
\includegraphics[width= 0.45\textwidth ]{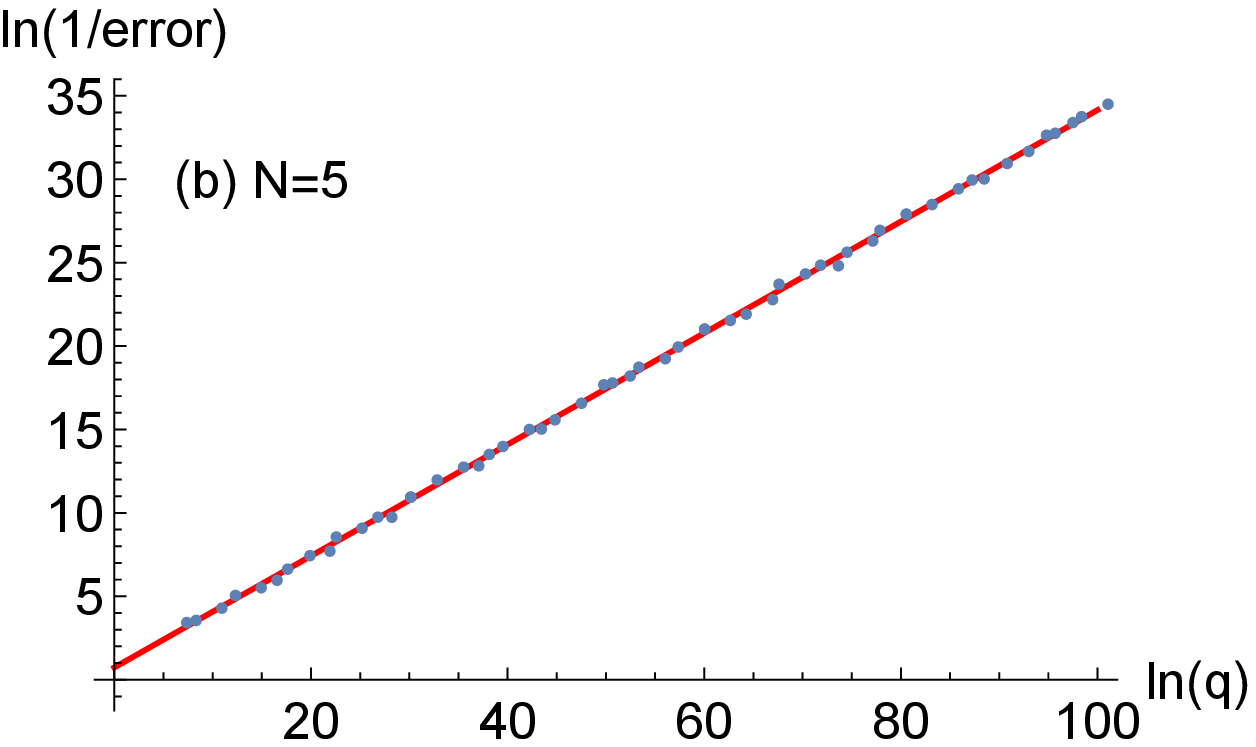}
\caption{Scaling relation between the number $q$ and the recurrence error $error $ defined in (\ref{recerror}). The straight lines are least squares approximations. In (a), $N = 15$, the slope of the line is $0.070846 \simeq 1/(N-1) = 0.071429 $; while in (b), $N = 5$, the slope is $0.334172 \simeq 1/(N-2) = 0.333333 $. In the latter case, the slope is not $1/(N-1)$ because of the integral relation $\sin (5\pi/12) = \sin(\pi/12) + \sin(3\pi/12)$.  }
\label{scaling}
\end{figure*}

Now for each $Q$ we can get a $q$ and a corresponding $error $. By taking $Q$ larger and larger, we get a series of pairs of $(q, error)$. It is necessary and straightforward to check the scaling relation between them. This we do in Fig.~\ref{scaling}. In Fig.~\ref{scaling}(a), we have $N = 15$, and $q$ scales with $error  $ as $q \propto error^{-14}$. However, in Fig.~\ref{scaling}(b), we have $N = 5$, and $q$ scales with $error  $ not as $q \propto error^{-4}$ but as $q \propto error^{-3}$. The reason is simply that for $N = 5$, the natural frequencies (\ref{frequency}) of the normal modes are not linearly independent over the integers. To be specific, as can be easily verified, we have the equality,
\begin{eqnarray}
  \omega_5 = 2 \omega \sin \frac{5\pi}{12} = 2 \omega \sin \frac{\pi}{12} + 2 \omega \sin \frac{2\pi}{12} = \omega_1 + \omega_3 .
\end{eqnarray}
Hence, for an arbitrary integer $q$, $q = q \alpha_1 + q \alpha_3$, which means, whenever $\langle q  \alpha_1 \rangle   $ is less than $\epsilon$, so is $\langle q  \alpha_3 \rangle  $ automatically. Therefore, the problem of looking for a $q$ such that
\begin{eqnarray}
  \max\{\langle  q \alpha_1 \rangle , \langle q \alpha_2 \rangle , \langle q \alpha_3 \rangle ,\langle q \alpha_4 \rangle\} \leq \epsilon
\end{eqnarray}
is actually equivalent to the problem of looking for a $q$ such that
\begin{eqnarray}
  \max\{\langle  q \alpha_1 \rangle , \langle q \alpha_2 \rangle , \langle q \alpha_4 \rangle \} \leq \epsilon .
\end{eqnarray}
This is why we have the $q\propto error^{-3}$ law.

\section{Discussions}

We have had the privilege to witness a Poincar\'{e} recurrence, which is a very improbable event---much rarer than the recurrence of Halley's comet for any realistic system actually. While this seems a hopeless task for a generic system,  it is possible in the case of a completely integrable system, thanks to the fact that the problem can be reduced to the classic Diophantine approximation problem, for which mathematicians have already prepared many theorems, and in particular, algorithms, for us. We have thus here a case in which number theory plays a vital role in solving physical problems. Note that while analysis and algebra are routinely used in physics nowadays, number theory presents itself rarely in physics.


But frankly speaking, we have achieved only partial success. While the LLL algorithm is powerful, its insufficiency is also apparent. For given $\epsilon$ and for a generic set of numbers $\{\alpha_{1\leq i \leq n } \}$, there are infinitely many $q$ satisfying (\ref{diophantine2}). Actually, collecting the appropriate $q$'s in a sequence $\{q_1, q_2, \ldots, \}$,  Weyl's theorem states that\cite{stein}
\begin{eqnarray}
  \lim_{m\rightarrow \infty } \frac{q_m }{m } &=& \frac{1}{\epsilon^n } .
\end{eqnarray}
That is, on average, the distance between two adjacent $q$'s is $1/\epsilon^n $. Physically, it means that recurrences with a precision on the order of $\epsilon $ will appear with a frequency on the order of $\epsilon^n $. However, the algorithm can get only one $q$ for each $\epsilon$. How to capture all the $q$'s is thus a problem of interest. As far as we know, this problem has been completely solved only for $n=1$.\cite{kovacs} For larger $n$, only partial success has been achieved. Moreover, the $q$ returned by the algorithm is often not the smallest one, i.e., it is not smaller than $1/\epsilon^n$, the bound given by the Dirichlet theorem, although it is on the same order.

Historically, Poincar\'e recurrence was discovered in the context of classical mechanics. However, it can be generalized to quantum mechanics straightforwardly.\cite{loinger, schulman, wallace} For the sake of simplicity, let us assume that the quantum system has a finite-dimensional Hilbert space, and the eigenstates and eigenvalues are $\{|m \rangle \}$ and $\{E_m \}$, respectively. The quantum Poincar\'e recurrence theorem then states that for any initial state
\begin{eqnarray}
  |\psi_0 \rangle  &=& \sum_{m=1}^N a_m |m \rangle,
\end{eqnarray}
the system, evolving as
 \begin{eqnarray}\label{evol}
  |\psi (t)\rangle  &=& \sum_{m=1}^N a_m  e^{-i E_m t } |m \rangle,
\end{eqnarray}
 will come back close to $|\psi_0 \rangle $ up to an arbitrary precision infinitely many times. Specifically, for any $t> 0 $ and $\epsilon > 0$, there exists a $t_r > t $ such that
\begin{eqnarray}\label{dist}
 d(|\psi(t_r)\rangle, | \psi_0 \rangle ) \equiv  ||\psi(t_r)\rangle - |\psi_0 \rangle | &\leq & \epsilon .
\end{eqnarray}
This is not surprising. As (\ref{evol}) shows, the wave function also evolves on an $N$-torus, and hence the formalism developed for a completely integrable system applies too. Quantitatively, we have
\begin{eqnarray}
 d^2(|\psi(t_r)\rangle, | \psi_0 \rangle )  &=& \sum_{m=1}^N |a_m|^2 (2- 2 \cos E_m t ) \nonumber \\
   &=& \sum_{m=1}^N |a_m|^2 4 \sin^2 \frac{E_m t }{2}.
\end{eqnarray}
For (\ref{dist}) to be satisfied for a generic initial state $|\psi_0 \rangle $, a sufficient condition is
\begin{eqnarray}\label{diophantine3}
  \langle  \frac{E_m t}{2 \pi } \rangle   &\leq & \frac{\epsilon}{2 \pi }
\end{eqnarray}
for all $ 1\leq m \leq N  $. Let $t = 2\pi q /E_N  $ with $q$ a whole number, so that $\langle  E_N  t /2\pi \rangle  = 0 $ and (\ref{diophantine3}) is satisfied already for $m = N $, we get the problem defined in (\ref{diophantine1}) again, with $\alpha_i = E_i /E_N$. Therefore, the same algorithm applies for a quantum system too.


\section*{ACKNOWLEDGMENTS}
We are grateful to Yang Ji and Hai-feng Song for helpful comments. This work is supported by the Fujian Provincial Science Foundation under Grant No. 2016J05004 and the Science Challenge Project under Grant No. JCKY2016212A502.

\section*{Appendix}
Here, we append the Mathematica code used to find the particular number in (\ref{particular}):
\begin{verbatim}
num = 15;
omega = 2 * Sin[Range[num]*Pi/2/(num + 1)];
alpha = Take[omega, {1, num - 1}]/omega[[num]];
Q = 10^35;
B = Table[0, {i, 1,  num}, {j, 1, num}];
B[[1, 1]] = 1;
For[i = 1, i <= num - 1, i++, B[[1, i + 1]] =
    Round[Q*alpha[[i]]];];
For[i = 2, i <= num, i++, B[[i, i]] = Q; ];
RB = LatticeReduce[B];
q = Abs[RB[[1, 1]]]
error = Abs[N[q*alpha - Round[q* alpha], 200]];
\end{verbatim}
Once the most crucial number $q$ is obtained, all other calculations can be done straightforwardly.

The readers can modify the code a little bit to reproduce the particular number in Eq.~(\ref{num1}).


\begin{thebibliography}{99}
\bibitem{non}
In other words, the motion of the system should be bounded. That this is necessary can be seen from the simple fact that a free particle in one dimension with a finite moment will never return to its initial state, neither classically nor quantum mechanically.
\end{thebibliography}
\end{document}